\documentclass[sigconf]{acmart}
\usepackage{color,soul}
\usepackage{wrapfig}
\usepackage{microtype}
\AtBeginDocument{%
  \providecommand\BibTeX{{%
    \normalfont B\kern-0.5em{\scshape i\kern-0.25em b}\kern-0.8em\TeX}}}

\setcopyright{acmcopyright}
\copyrightyear{2021}
\acmYear{2021}
\acmDOI{10.1145/1122445.1122456}

\acmConference[TechDebt '21]{TechDebt '21: 4th International Conference on Technical Debt}{May 19--21, 2021}{Madrid, Spain}
\acmBooktitle{TechDebt '21: 4th International Conference on Technical Debt, May 19--21, 2021, Madrid, Spain}
\acmPrice{15.00}
\acmISBN{978-1-4503-XXXX-X/18/06}

\begin{document}

\title{The Need for Holistic Technical Debt Management across the Value Stream: Lessons Learnt and Open Challenges}

\author{Somayeh Malakuti, Jens Heuschkel}
\email{{somayeh.malakuti, jens.heuschkel}@de.abb.com}
\affiliation{%
  \institution{ABB AG Corporate Research Center, Germany}
}


\renewcommand{\shortauthors}{Malakuti, et al.}

\begin{abstract}
The long lifetime and the evolving nature of industrial products make them subject to technical debt at 
different levels. Despite multiple years of research on technical debt management, our industrial experience shows that 
introducing systematic technical debt management in a large-scale company is very challenging.
To identify the challenges, we provide a conceptual framework for holistic debt management across the product development value stream, which takes 
multiple categories of debt and their interplays into account. We use this framework to identify multiple challenges that are still open to be explored by the research community. 
Due to the practical nature of technical debt management,
we believe this paper can guide the research community on the needs of industry for the effective application of technical debt management in practice.
\end{abstract}




\maketitle

\section{Introduction}

Technical debt can easily be observed  in industrial products and systems (e.g., controllers, smart sensors, robots), 
which have a very long lifetime and their features evolve over time. These products have an interdisciplinary nature, as 
they consist of mechanical, electrical and software components. 
Introducing systematic technical debt management in a company is ultimately about increasing the maturity level of the company processes to include technical debt management. However, our experience shows that this cannot be done in isolation, without considering the overall maturity level of the company in its product development value stream. 

In the literature, there are several studies \cite{LI2015193,Potdar2020,BESKER20181} that focus on certain technical debt management techniques at code level, software architecture level and/or across software/electrical/mechanical components; these studies assume certain level of infrastructure, methods, and competence maturity 
in companies to apply these techniques in practice.
There are studies \cite{MARTINI201842,Neil2015,YLIHUUMO2016195} assessing the maturity level of companies in this regard, 
without providing details on challenges and guidelines on moving from one maturity level to another one in practice.
 There are studies \cite{HOLVITIE2018141} confirming the positive impacts of the agile processes and practices 
(e.g., backlogs, retrospective) in better technical debt management, without considering the negative 
impacts of agile processes if not properly implemented across the value stream of product development. 

Based on our industrial experience in a large-scale multi-national company \cite{malakuti2020}, we observe that 
the above-mentioned gap in the technical debt literature leaves practitioners with multiple open challenges 
to increase the maturity level 
of companies in their technical debt management.  Although the need for looking at technical debt at a more holistic level beyond software code and architecture has been observed in the literature \cite{Neil2015,YLIHUUMO2016195,RIOS2018117,TOM20131498},  no conceptual framework for holistic debt management as the base for identifying open challenges has been proposed in the literature. 

The BAPO model \cite{bapo} conceptualizes different levels of alignment among Business, Architecture, Process and Organization. We take this model as the reference to further detail the need for 
taking a holistic perspective on the topic of technical debt management across the value stream of the product development. 
Such holistic debt management takes various kinds of debt such as technical, process, people, infrastructure, and portfolio debt into account.
In addition to individual kinds of debt, it is mandatory to consider the relations among these kinds of debt as well.
This leads to a complex graph of interrelated debts, which cannot be effectively managed if one only focuses on one category of debt. Managing multiple categories of debt and their relations require suitable methods to formally model and reason about them; such methods have not been studied sufficiently in the literature. 

Based on our experience, the lack of a holistic approach for debt management leads to 
reactive management of debt by shifting the focus from one category to another, causing even more debt through further sub-optimal decisions.   

As for the contirbutions, we explain our industrial case study, and the challenges that we observed in applying technical debt management in practice.
Then, we sketch a conceptual model to depict various aspects and elements of a holistic debt management method.
Finally, we outline multiple open challenges that could be taken by the research community in future.

\section{Industrial Experience Overview}
Our experience is based on a project in the form of external consultancy that we have been offering to 
an industrial company as our client
since 2018 in multiple phases. The industrial products manufactured by the company have mechanical, electrical and software components.
The company historically has stronger background in the mechanical and electrical components, 
but  software has been gaining strong importance over the time as well.

Throughout the project, eight engineers, 
one process manager and one technology manager closely participated in the case study.
The participants have strong background in hardware design, while
software engineering competence differed among them.
We have adopted different approaches such as frequent teleconferences, face to face workshops, field observation, 
and architecture co-design. We have performed multiple interviews with managers and engineers on their perception on technical debt, 
as well as a survey with 100 participants to acquire their perspective on this topic. More details can be found in \cite{malakuti2020}.

Before the start of the project in 2018, there were several internal workshops 
in the company with the aim of achieving bug-free software. 
Those workshops concluded that insufficient modularity of the software is the main reason for 
the increasing number of bugs as well as high maintenance effort. 
As a result, the project started in 2018 with the aim of validating 
this hypothesis. Our activities throughout the project are summarized as follows:

\vspace{-0.7em}
\paragraph{Phase 1: Assessing the modularity status of the software:}
The phase 1 of the project was about getting to know the software better and validating the hypothesis of our client 
regarding the low modularity of their software. We identified various kinds of modularity-related technical debt that exist at the architectural and code levels.

\vspace{-0.7em}
\paragraph{Phase 2: Prioritizing and repaying technical debt at the code level:} 
Refactoring large-scale legacy software as a whole requires significant amount of time and resources, 
which could not be offered by our client. 
In the phase 2, a high-level architecture sketch was defined by our client, and we were requested to accompany developers with the 
refactoring of two pilot modules based 
on the sketched high-level architecture.

Although certain refactoring was applied, we still did not manage to fully reach the goals of this phase.  
Inadequate specification of architecturally-significant requirements and concerns, 
and accordingly inadequate architectural technical debt repayment by our client engineers were among the reasons that we 
could not reach our refactoring goals. Moreover, we observed that in addition to architecture and code debt, 
there is debt in test cases and build scripts. The lack of clear specification of relations 
among different kinds of debt prevented us to manage them systematically. 
 
\vspace{-0.7em}
\paragraph{Phase 3: Adopting a systematic technical debt management approach:}
As it is studied, architectural debt can be the major source of technical debt \cite{Neil2015}. Based on our learnings from the phase 2, 
in the phase 3 we aimed at iteratively improving the architecture and repay debt by adopting the state-of-the-art 
methods for identifying and prioritizing architectural technical debt \cite{almeida2018aligning,Lenarduzzi2019TechnicalDP}. 
In addition, we integrated architectural metrics offered by the tools Lattix \cite{lattix} and SonarQube \cite{sonarqube} (beside other) 
into a common dashboard so that engineers can monitor the quality trend of software.
 The phase 3 led to the following learnings and results:

a) Lack of common understanding of various fundamental concepts such as 
architecture-centric software development, quality attributes, quality scenarios, 
and trade-off analysis significantly impeded the progress of the work. Therefore, our focus was shifted from repaying architecture debt to repaying people debt by supporting the engineers in gaining deeper insight on these topics by offering some trainings, and planning for more trainings. 

b) Agile principles require software teams to be empowered to perform different tasks. 
Naturally, in a large-scale company with several hundred developers, 
time and budget constraints would not allow harmonizing 
competences of everyone at once. 
As part of defining the strategies for repaying the people debt, there were different opinions on whether it is feasible for everyone to receive training or 
whether the (temporary) role of 'software architect' should be defined in teams. Due to its organizational impacts, this discussion is still open. 

c) In addition to the people debt, 
we observed the need for repaying some process debt to 
mandate the engineers to perform quality attribute workshops, architecture trade-off analysis, etc.

d) The introduction of Lattix \cite{lattix} 
and its quality metrics laid the foundation for objective assessment of the architecture. 
Although more source code analysis tools could also be used to enrich the set of calculated metrics, this idea did not get traction in the company.
The main reason was that the introduction of each new tool had be accompanied by supporting the 
engineers to adopt the tool in their daily work. 
Otherwise, a new tool could eventually contribute to the infrastructure debt by making the infrastructure more complex while 
not being actively and consistently used.

e) We observed that if the activities to repay various kinds of debt (people debt, technology debt, architecture debt, etc.) were not well coordinated, they could put extra burden on the engineers to deal with the imposed deficiencies. For example, if parallel activities to establish a common understanding of software architecture topics, to introduce Lattix in the infrastructure, and to introduce the software product line technology in some parts of the system were not well coordinated and prioritized, they could prevent the engineers to master these topics and apply them in practice effectively. 

Overall, our experience showed that increasing the maturity level of a company in its technical debt management approach is a gradual process; however, the steps of this process are not studied in the literature to guide us through. Moreover, 
one cannot introduce a systematic technical debt management approach in a company without taking a holisitic approach that takes various categories of debt and their relations into account.

\vspace{-1em}
\section{Towards Holistic Debt Management}
 \begin{figure*}[t!]
 \centering
 \includegraphics[width=1\textwidth]{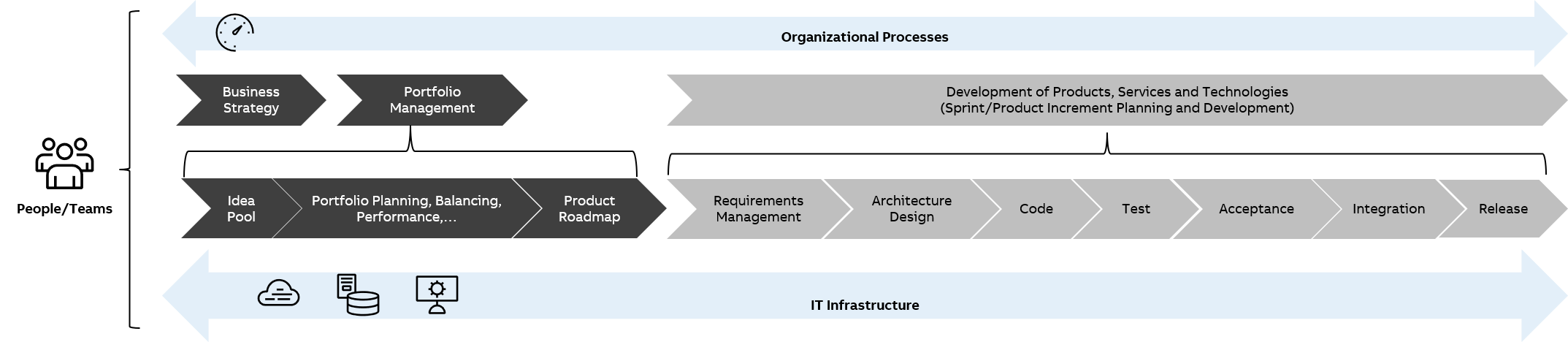}
 \caption{An example value stream of system development.}
 \label{fig:valuestream}
 \end{figure*}

Among others, one reason that technical debt occurs in systems is the strong demand for business agility, 
which creates time and cost pressure during the development of the systems. This may lead to the assumption that companies 
are doing well in terms of their agile processes, and need to dedicate more focus on technical debt management to make sure these 
two aspects remain balanced. However, this may not be a correct assumption always. 
To clarify this, let us look at an example value stream of system development, illustrated in Figure \ref{fig:valuestream}.

 
The value stream starts by defining the business strategies of the company for its products.
Portfolio management is an important step, which contains preparing the idea pool, portfolio planning and balancing, 
and defining the products roadmap. The products roadmap is input to the requirements management phase, 
which focuses on soliciting and clarifying customer requirements.
This phase is followed by other development phases such as architecture design, implementation, test and release. 

There are various organizational processes that define workflows and methods to perform tasks in each phase. 
Different phases of the value stream require different people competences; 
The phases make use of some IT tools and infrastructure, which are for example software development tools, 
build tools, databases and file systems to store the artifacts.

As shown by Leopold et.al. \cite{agile}, companies may not get the desired throughput 
even though in their estimate they adopt agile processes. Three reasons have been identified for this:
a) no agile interactions among teams that work on the same products, leading to unmanaged dependencies among product features and teams,
b) no end-to-end management of the value stream, for example, by emphasizing more on the agility in development phases 
while ignoring the delays in earlier phases of the value stream,
and c) no agile strategic portfolio management, causing a large number of initiatives overloading teams.

 \begin{figure}[b!]
 \centering
 \includegraphics[width=0.4\textwidth]{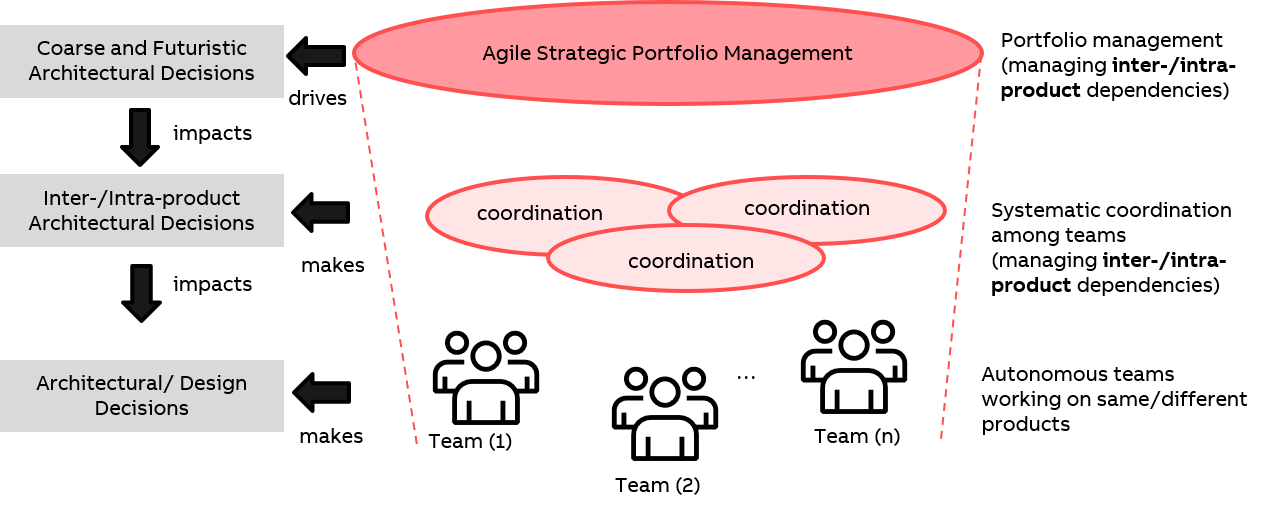}
 \caption{Agile processes and technical debt}
 \label{fig:teams}
 \end{figure}

These aspects in the proper adoption of agile practices (i.e. process debt) cause time pressure, 
which can eventually cause various technical debt in the systems.
Besides the time pressure, our experience shows that these problems can cause technical debt in other ways too.
Consider Figure \ref{fig:teams} for example. 
At the team level, we have autonomous scrum teams that make various design decisions for individual products or 
sub-products, which may also introduce architectural, code, test and/or build debt.

 
At the inter-team level, we require systematic coordination among these teams to make coarse-grained architectural decisions for managing the inter- and intra-product dependencies. 
Insufficient means and guidelines for such systematic coordination (i.e. process debt) will lead to unmanaged architectural decisions (i.e. architectural debt).
At the portfolio level, we require agile portfolio management that takes the products interdependencies into account, 
so that more coarse-grained and futuristic architectural decisions can be taken in a systematic way.
Lack of such portfolio management approaches (i.e., portfolio debt) will also lead to ad-hoc architectural decisions (i.e. architectural debt).

\begin{figure*}[t!]
 \centering
 \includegraphics[width=1\textwidth]{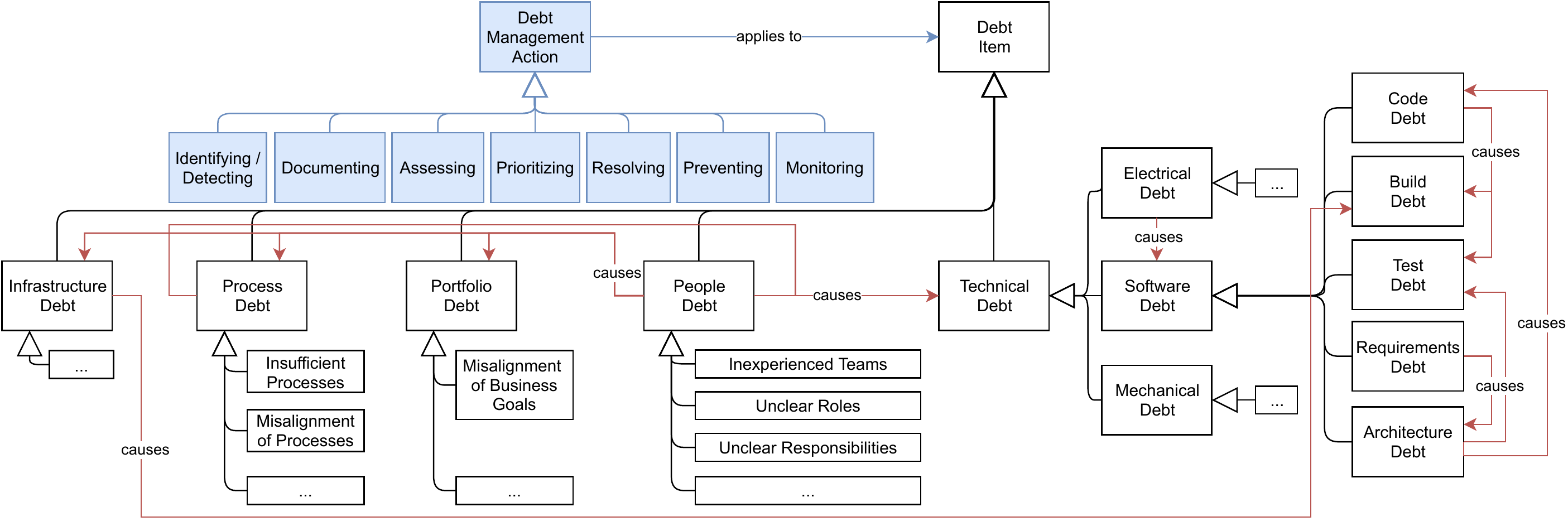}
 \caption{Scope of holistic debt, observed relations, and holistic debt management.}
 \label{fig:debt-uml}
 \end{figure*}

This example shows that sub-optimal decisions in one part of the value stream may propagate within or across 
other parts causing debts at those parts. Therefore, 
to effectively manage technical debt in large-scale companies, we need to proactively identify debt at each part, identify 
the cause-effect relations among them and manage them accordingly. This indicates the need for 
a holistic debt management approach, which enables us to move from reactive decision making to proactive and conscious 
decision making for debt management across the value stream. 
Figure \ref{fig:debt-uml}
depicts various aspects of such a holistic approach. 


For each kind of debt, the systematic debt management actions that include 'identifying/detecting debt', 'assessing the debt impacts', 'prioritizing the debt', 'resolving the debt', 'preventing further debt', 'documenting debt', 
'monitoring debt' should be applied. 

In addition to systematically managing each category of debt, 
the relations among debts within and across each category should 
be identified and managed. For example in our case study, the process debt due to insufficient cross-team coordination, also the portfolio debt due to unmanaged inter-product relations have led
to some architectural debt; for example, product-specific features were not well modularized from cross-product features.
This debt was reflected in the software code, leading to some large classes with excessive amount 
of '\#if defined' statements to select the right product feature based on some build parameters.
Insufficient modularity of the code has led to insufficient unit tests as well as not well-modularized build scripts. 
There is also infrastructure debt due to scattered build scripts across repositories, which makes the update of build scripts even more difficult.

In addition to the above-mentioned debt in software components, some debts in electrical components 
have forced software developers to provide extra checks in the code to deal with the consequences of the debts in the electrical components. 
Those checks do not belong to the core functionality of the software components. 


The impacts of such cause-effect relations must be assessed, and be taken into account while prioritizing debts and while 
defining the order in which the debts must be repaid. Otherwise, as experienced in the phases 2 and 3 of our project, 
we may need to reactively shift the focus from one category of debt to another category, 
while losing the allocated time for debt management or even introducing more debt due to sub-optimal decisions made along the way.
\section{Open Challenges}
The extent of work depicted in Figure \ref{fig:debt-uml} 
is rather large, and naturally it cannot be applied at once.
Here, the major challenge that we observe is to define a roadmap that specifies the right combination and the right 
order of activities to effectively manage and repay various kinds of debt over time. 
For this we face several challenges and open questions.

\vspace{-0.7em}
\paragraph {A taxonomy of debt categories and their relations:} 
The starting point for holistic debt management is to have clear definition of different debt categories and possible relations among them. Although various taxonomies and conceptual models of technical debt exist \cite{RIOS2018117,TOM20131498}, our observation is that they are not consistent and more importantly do not consider various relations of debt categories with each other. We have sketched an example cause-effect relation in Figure \ref{fig:debt-uml} based on our case studies
. However, we need more extensive studies across industries to reach a comprehensive model. 

\vspace{-0.7em}
\paragraph {Systematic management of each debt category:}
Informally, one may consider (technical) debt as the delta between the 
current state and the desired target state in terms of various quality attributes. Hence, 
to manage debt we need to define the desired target state and its desired qualities, measure the delta, 
and define strategies to move from the current state to the target state. 
For each category of debt it should also be studied how the steps of systematic dept management depicted in
Figure \ref{fig:debt-uml} 
can be applied in practice. 

Due to the longevity of research work on code and architecture smells, 
concrete examples of code and architecture debt exist in the literature, and to some extent there is tool support 
to identify and assess them \cite{LI2015193,Potdar2020,BESKER20181}. 

For process improvement, CMMI \cite{cmm} lays the foundation for assessing the capability and maturity of processes. We opt to adopt this for identifying and resolving process debt. Here, we may consider process debt as 
the difference between the capability/maturity level of current processes 
and the desired target level. CMMI provides some guidelines and best practices to repay the process debt by increasing the capability/maturity level of the current processes. 
We believe that practitioners can benefit from similar guidelines and best practices for systematically managing other types
of debt. 

\vspace{-0.7em}
\paragraph {Systematic management of debt relations:}
Although the need for managing debt beyond code and architecture has been observed in the literature, for example in \cite{Quang2019,Klinger2011,RIOS2018117}, there is still noticeable need for systematic studies addressing different categories of debt and their relations.   

For example, the current studies on principle and interest of technical debt such as \cite{AMPATZOGLOU2020106391,Ampatzoglou2016} 
assume that some architectural technical debt has already be identified, and demand architects to provide an estimation of interest if the payment of debt is postponed.
However, as we have discussed earlier, the identification of technical debt itself 
requires enough competence and tools whose absence is presented as people debt and infrastructure debt. 
In addition, as depicted in Figure \ref{fig:debt-uml},
process debt and/or portfolio debt may be the 
root cause of some architectural debt. Therefore, the cost to identify, assess, prioritize and repay such 
debt as the cause of architectural debt must be considered as well; 
otherwise, the decision to repay architectural debt will remain a sub-optimal decision.

Where we cannot focus on one category of debt, we cannot fully repay multiple categories of debt at once. 
Hence, we need to define strategies to gradually repay some debt from some categories, and accept some debt to be paid later.
Figure \ref{fig:debt-uml}
is rather abstract and does not detail specific examples of each debt category and their relations. If all these details are identified, then prioritization of debt can no longer be effective without suitable means for formally modelling the causal relations of different debt, and without adopting suitable multi-objective optimization approaches. We believe industry can significantly benefit from studies on effective ways of defining iterations to gradually repay various categories of debt. 
  
\vspace{-0.7em}
\paragraph {Integrated debt management processes:}
Technical debt management and more generally holistic debt management are ultimately processes, 
which should be incorporated into the existing processes of companies.
There are studies \cite{MARTINI201842,YLIHUUMO2016195} that define maturity levels of technical debt management in companies, however, 
without providing specific guidelines 
on how to increase the maturity level, and without studying the interplays of technical debt management process with other product 
development processes. There are studies \cite{HOLVITIE2018141} assessing the impacts of agile processes and practices on technical debt management, concluding the positive impacts of agile practices such as code reviews. However, we identified that to assess the impacts of agile processes on technical debt management, one needs to assess the adequate realization of agile processes across the entire value stream.

We believe that the above-mentioned gaps in existing studies must be filled to be able to effectively increase the 
maturity level of companies in their holistic debt management. 

\vspace{-0.7em}
\paragraph {Insufficient objective measures to fill the communication gap}
The management of debt across the value stream requires  
communication among different stakeholders, which should be supported by objective metrics. It has been claimed that the notion of 'technical debt management' can fill the communication gap between the 
technical teams and the management teams because of its emphasize on both technical and financial aspects. However, our observation is that although extensive work has been performed on code and architectural metrics, the existing metrics and measures may not suffice even for the communication among the technical team, for example to predict the impacts and growth rate of technical debt. 

Although the need for more objective measures has been observed in the literature \cite{HOLVITIE2018141,Neil2015}, we miss systematic studies on catalogue of suitable metrics for different stakeholders within and across debt categories, as well as means to calculate and validate them.
From the process perspective, CMMI levels 4 and 5 expect that an organization reaches the level of data-driven decision making and self-optimization. This requires the identification of the stakeholders and their desired metrics, the identification of relevant data for measuring each category of debt, means to collect and cleanse the data, means to identify and model correlations among data, and suitable algorithms to analyze and reason about trends of debt. Our experience shows that companies usually have to develop their in-house solutions for this matter.  Here, we see noticeable potential for research on metrics and tools for assessing the impacts and growth rate of different debt categories individually and at holistic level, for different kinds of stakeholders. 
In this context,  an interesting follow up question might be in how far one can weave the debt types into the CMMI model were CMMI potentially falls short (e.g. people debt).

\section{Conclusions}
Based on our experience in introducing systematic technical 
debt management in an industrial company, we explained that technical debt management boils down to increasing the maturity level of a company across the value stream of its products development. Although the need for holistic approach for technical debt management has been observed in the literature, there was no proposal to concretize this topic. We took a step towards this by outlining a conceptual model for holistic debt management, which depicts various categories of debt and their relations. Based on this model, we discussed that some areas have been receiving stronger focus from research community than others. Consequently, insufficient focus on the holistic aspects of debt management eventually becomes a barrier to apply existing technical debt management approaches in practice. We believe that this paper can help the research community in defining further research topics in the area of technical debt management.  

\bibliographystyle{ACM-Reference-Format}
\bibliography{literature}

\end{document}